
\input manumac
\twelvepoint

\def\M{{\cal M}}

\line{\hfil DAMTP/R-93/19}

\vskip 0.4truecm
\centerline{\bf Vacuum interpolation in supergravity via super p-branes}
\vskip 1truecm
\centerline{G.W. Gibbons and P.K. Townsend}
\vskip 1 cm
\centerline{DAMTP, Univ. of Cambridge,}
\centerline{Silver St., Cambridge, U.K.}

\bigskip

\centerline{\bf ABSTRACT}
\vskip 0.5cm
We show that many of the recently proposed supersymmetric p-brane
solutions of d=10 and d=11 supergravity have the property that they
interpolate between Minkowski spacetime and a compactified spacetime,
both being supersymmetric supergravity vacua. Our results imply that the
effective worldvolume action for small fluctuations of the super p-brane is a
supersingleton field theory for $(adS)_{p+2}$, as has been often conjectured in
the past.

\vfill\eject

It is possible that particle physics in our four-dimensional (d=4) universe
may ultimately be well-described by some compactification of a ten-dimensional
(d=10) supergravity theory that serves as the effective field theory of a d=10
superstring theory. Even if superstring theory meets with complete success in
this respect there will remain the question of why the universe `chooses' to
compactify six dimensions in a particular way and, indeed, why it `chooses' to
compactify any of them since d=10 Minkowski spacetime ($\M_{10}$) is as good a
vacuum solution as any other from a purely mathematical point of view. In
contrast to solutions of simple flat space field theories there is no way
to compare the energies of different compactifications and thus determine `the'
vacuum by finding the one of lowest energy. In these circumstances it might be
supposed that the choice of compactification must be left to some theory of
initial conditions. An alternative is that {\sl all possible} compactifications
are already to be found in different spatial regions of a single (presumably
ten-dimensional) universe. The particular region in which we find ourselves
must then be decided by chance and/or anthropic considerations. Ideas along
these lines, but within the context of a four-dimensional universe, have been
suggested previously by Linde [1], and the possibility of an interpolation
between different compactifications of d=11 supergravity was suggested by van
Baal et al. in their work on `local compactification' [2].

A clue to progress in this direction is provided by consideration of the
extreme
Reissner-Nordstrom (RN) black hole as a solution of N=2 d=4 supergravity. This
solution interpolates between 4-dimensional Minkowski spacetime ($\M_4$), at
spatial infinity, and $(adS)_2\times S^2$, down an infinite wormhole throat
[3]. Both asymptotic spacetimes are maximally-supersymmetric `vacua' of N=2
supergravity. We shall show here that many of the recently discussed extreme
black p-brane solutions of d=10 [4-7] and d=11 [8,9] supergravity also
interpolate between supersymmetric (although not always maximally
supersymmetric) vacua. The cases that most closely resemble the RN prototype
are

\medskip
(i) d=11 membrane  (p=2)
\medskip
(ii) d=11 fivebrane  (p=5)
\medskip
(iii) d=10 IIB self-dual threebrane  (p=3)
\medskip
For these cases the p-brane interpolates between $\M_d$ and $(adS)_{p+2}\times
S^{d-p-2}$. The latter spacetimes are known to be maximally supersymmetric
solutions (for the appropriate value of p) of the respective supergravity
theories [10-12]. Like the extreme RN black hole, these p-brane solutions are
non-singular and break only half the supersymmetry; they may therefore be
regarded as examples of `supersymmetric extended solitons'.

An example that does not quite fit the above pattern is the
\medskip
(iv) d=10 fivebrane
\medskip\noindent
We shall show that this solution interpolates between $\M_{10}$ and
$\M_7\times S^3$. This is something of a surprise since no compactification of
d=10 supergravity to $\M_7$ on $S^3$ has been previously described. As we
shall see, the explanation lies in the asymptotic behaviour of the dilaton
field
down the wormhole throat; rather than approach a constant, as it does for cases
(i)-(iii) above, it approaches a linear function of the inertial
coordinates of $M_7$ (this behaviour is similar to certain d=4 `dilaton black
holes' [13]). We shall show that there is indeed such a compactification of
d=10
supergravity but it is not maximally supersymmetric because, like the fivebrane
solution itself, it breaks half the supersymmetry. That is, unlike cases
(i)-(iii), the full supersymmetry is not restored in {\sl both} asymptotic
regions for case (iv). Nevertheless, the fivebrane solution is non-singular and
supersymmetric, so it too can be regarded as a `supersymmetric extended
soliton'.

Black fourbrane and sixbrane solutions of d=10 supergravity were also
described in [6]. The extremal cases of these solutions were shown in [14] to
break only half the supersymmetry and, on this basis, were proposed as
candidates for further `supersymmetric extended solitons'. However,
these solutions are singular in the extremal limit and do {\sl not}
interpolate between solutions of d=10 supergravity.

A further implication of our work concerns the nature of the effective
worldvolume action for the p-brane solutions of cases (i)-(iii) in which there
is an interpolation between Minkowski spacetime and a lower-dimensional anti
de Sitter spacetime. Far down the wormhole throat we have a supergravity
theory compactified on a (d-p-2)-sphere to $adS_{p+2}$. It is known from
studies of these compactifications that the fields of a singleton
supermultiplet of the adS supergroup appear as coefficients in the harmonic
expansion of the d-dimensional fields on the (d-p-2)-sphere, but that these
can be gauged away everywhere except at the boundary of adS [15]. This result
is
in accord with the currently accepted field theoretic interpretation of
singleton irreps of adS groups, ie. that they are what we would now call
topological field theories in that all physical degrees of freedom reside on
the boundary [16-18]. In the current context the boundary is just the opening
of
the wormhole throat, which is perceived from the exterior as the p-brane core .
We therefore conclude that the worldvolume fields of the effective p-brane
action should be those of the appropriate adS singleton supermultiplet. These
are as follows [19]:
\medskip
(i) 3-dimensional N=8 scalar supermultiplet
\medskip
(ii) 6-dimensional N=2 antisymmetric tensor multiplet
\medskip
(iii) 4-dimensional N=4 Maxwell supermutiplet
\medskip
In cases (i) and (iii) it is known that these are indeed the worldvolume fields
by an analysis of the fluctuations about the p-brane solution [5,20]. In case
(ii) one can deduce the worldvolume fields from the fact that they must
include five Goldstone scalars associated with the breakdown of translation
invariance at the p-brane in the five transverse directions and the fact that
these fields must appear in an N=4 worldvolume supermultiplet, because this
corresponds to the correct number of supersymmetries left unbroken by the
solution. The unique 6-dimensional N=4 supermultiplet with five scalars is the
antisymmetric tensor multiplet [21]. Note that the fermions of this multiplet
transform as a 4-plet of ${\rm USp}(4)\sim {\rm Spin}(5)$, as expected since
an $SO(5)$ group of rotations in the five-dimensional `transverse' space is
left unbroken by the fivebrane solution.

It has been suggested at various times in the past that effective actions
for super p-branes might be considered as supersingleton field theories
associated with the appropriate supergroup extension of the $(adS)_{p+2}$
group [22,23,19]. Since this group acts as a conformal supergroup on the
boundary and since the superstring action has worldsheet conformal-invariance
(in the `conformal' gauge) this proposal is most natural for superstrings.
However, it is known that the effective action for {\sl small} (i.e. to
quadratic order) fluctuations of a membrane at the boundary of adS space is
also conformally invariant [19,20]. Our results provide significant further
evidence for the connection between supersingleton field theories and super
p-branes.

All of the metrics to be considered here have the form
$$
\eqalign{
ds^2_{d} =& -\big[ 1-\big({r_+\over r}\big)^{(d-p-3)}\big]
\big[ 1-\big({r_-\over r}\big)^{(d-p-3)}\big]^{\gamma_x -1} dt^2 +
\big[ 1-\big({r_-\over r}\big)^{(d-p-3)}\big]^{\gamma_x} d{\bf x}\cdot d{\bf
x}\cr & +\big[ 1-\big({r_+\over r}\big)^{(d-p-3)}\big]^{-1}
\big[ 1-\big({r_-\over r}\big)^{(d-p-3)}\big]^{\gamma_r} dr^2 \cr
&+ r^2 \big[ 1-\big({r_-\over r}\big)^{(d-p-3)}\big]^{\gamma_r +1}
d\Omega^2_{(d-p-2)} }
\eqno (1)
$$
for constants $\gamma_x$ and $\gamma_r$, and where $d\Omega^2_{(d-p-2)}$
is the `round' metric on the (d-p-2)-sphere (although it is
straightforward to extend the results we will obtain below to the case in
which it is any Einstein metric on $S^{(d-p-2)}$). These metrics are asymptotic
to the flat metric on d-dimensional Minkowski space, $\M_d$, as $r\to \infty$.
They have an outer horizon at $r=r_+$ and an inner horizon at $r=r_-$, in close
analogy with the RN black hole solution of d=4 Einstein-Maxwell theory, which
is
in fact a special case of the above metric with $d=4$, $p=0$, $\gamma_r=-1$ and
$\gamma_x=2$. In those cases for which there is a dilaton field field $\phi$,
it will take the form
$$
e^{-2\phi} = \big[ 1-\big({r_-\over
r}\big)^{(d-p-3)}\big]^{\gamma_\phi}
\eqno (2)
$$
for some constant $\gamma_\phi$ (in the solutions of interest here). We are
principally interested in the extreme case for which $r_+=r_-=a$. In this case
the metric is $$
\eqalign{
ds^2_d = &\big[ 1-\big({a\over r}\big)^{(d-p-3)}\big]^{\gamma_x}\big[ -dt^2 +
d{\bf x}\cdot d{\bf x}\big] \cr
&+\big[ 1-\big({a\over r}\big)^{(d-p-3)}\big]^{\gamma_r -1}dr^2  + r^2
\big[ 1-\big({a\over r}\big)^{(d-p-3)}\big]^{\gamma_r +1} d\Omega^2_{d-p-2}}
\eqno (3)
$$
which can be interpreted as the metric of a flat, static, and infinite
p-brane. However, only if $\gamma_r=-1$ is this metric non-singular at
$r=a$, so only in this case are these solutions candidates for `extended
solitons'. Note that this excludes all but the threebrane case of those
metrics considered in [14]. The extremal p-branes of cases
(i)-(iv) mentioned above all have $\gamma_r=-1$ and are non-singular at $r=a$.
The other exponents are, in each of these cases,
\medskip
(i) $\gamma_x ={2\over3}$
\medskip
(ii) $\gamma_x = {1\over 3}$
\medskip
(iii) $\gamma_x={1\over2}$ and $\gamma_\phi =0$
\medskip
(iv) $\gamma_x=0$ and $\gamma_\phi =1$
\medskip

To examine the behaviour as $r\to a$ in these cases we return to the
non-extremal metric (1) and introduce the new coordinates
$$
t' = \big({(d-p-3)\delta\over a}\big)^{-{\gamma_x\over2}} t \qquad {\bf x}' =
\big({(d-p-3)\delta\over a}\big)^{-{\gamma_x\over2}}
 {\bf x} \qquad \rho= {a\over
d-p-3}{\rm ln}\big({r-a\over\delta}\big)\ ,
\eqno (4)
$$
where $\delta$ is defined by $\delta^2= r_+-r_-$, and we have now set $r_+=a$.
The horizon is now at $\rho=\infty$. For fixed $r$, $\rho\to \infty$ in the
extremal limit $\delta\to 0$, so the new coordinates are appropriate for a
description of the asymptotic metric down the wormhole throat. Taking the
$\delta\to 0$ limit we find that
$$
ds^2_d \to  e^{\gamma_x
(d-p-3)\rho\over a}[ -(dt')^2 +d{\bf x}'\cdot d{\bf x}'] + d\rho^2 + a^2
d\Omega^2_{d-p-2} \ .
\eqno (5)
$$
Similarly, we find the asymptotic form
$$
\phi = {\rm const} -\big({\gamma_\phi(d-p-3)\over a}\big) \rho
\eqno (6)
$$
of the dilaton (when applicable). When $\gamma_x\ne 0$ the metric (5) is an
invariant metric on $(adS)_{p+2}\times S^{d-p-2}$. This happens for cases
(i)-(iii) for which the corresponding p-brane solution interpolates between
$\M_d$ and
\medskip
(i) $(adS)_4\times S^7$
\medskip
(ii) $(adS)_7\times S^4$
\medskip
(iii) $(adS)_5\times S^5$
\medskip
\noindent
These are all known compactifications of d=11 and d=10 supergravity,
preserving all the supersymmetry.

The above three cases are all closely analogous to the d=4 extreme RN black
hole, but this analogy is closest for the self-dual three-brane as we now
explain. The extreme RN black hole has a conformal isometry that exchanges the
two asymptotic regions [24]. The generalization to p-brane solutions of the
form (1) with $\gamma_r=-1$ involves the consideration of a
new radial coordinate $\tilde r$ given by
$$
r^{(d-p-3)} -a^{(d-p-3)} = {a^{2(d-p-3)}\over {\tilde r}^{(d-p-3)}-
a^{(d-p-3)}   }
\eqno (7)
$$
The new metric is then conformal to the original one if $\gamma_x= 2/(d-p-3)$.
This condition is satisfied by the d=4 extreme RN solution. Of the above three
p-brane solutions it is satisfied only by the self-dual threebrane.

For the d=10 fivebrane we have instead an interpolation between $\M_{10}$ and
\medskip
(iv) $\M_7\times S^3$
\medskip
\noindent
with a dilaton that is linear in the inertial coordinates of
$\M_7$. Such a compactification of $d=10$ supergravity has not been previously
described in the literature but it must exist because a solution of the field
equations remains a solution in the limit $\delta\to 0$ discussed above.
To verify this we shall need the bosonic action of d=10 supergravity
$$
S=\int d^{10} x\sqrt{-g} e^{-2\phi}\big[ R + 4(\partial\phi)^2
-{1\over12}H^2\big]\ ,
\eqno (8)
$$
where $H$ is a three form field strength. The field equations are
$$
\eqalign{
0&=R_{MN} -{1\over4}H_{MPQ}H_N{}^{PQ} + 2\nabla_M\partial_N\phi \cr
0&= \nabla_M\big(\sqrt{-g} e^{-2\phi}H^{MNP}\big)\cr
0&= 4(\partial\phi)^2 -4\nabla^2\phi -R +{1\over12}H_{MNP}H^{MNP}\ . }
\eqno (9)
$$
We now split the coordinates $x^M$ into two sets, $x^\mu\; (\mu =0,1,\dots,6)$
for $\M_7$ and $y^m\; (m=1,2,3)$ for a compact 3-space, and we make the ansatz
$$
H_{mnp}= k e_m{}^a e_n{}^b e_p{}^c \varepsilon_{abc} \qquad \phi={1\over2}k'
n\cdot x
\eqno (10)
$$
where $e_m{}^a(y)$ is the dreibein for the 3-space, $k$ and $k'$ are constants,
and $n$ is a unit spacelike 7-vector. The Einstein equation now yields
$$
R_{ab}={1\over2}k^2\delta_{ab}
\eqno (11)
$$
which implies that the 3-space is $S^3$ with inverse radius $k$. The
antisymmetric tensor equation is trivially satisfied while the $\phi$ equation
is satisfied if $k'=k$. Hence $\M_7\times S^3$ {\sl with a linear dilaton} is
a solution.

We remark that a similar compactification to $\M_4$ on $S^3\times
S^3$ also exists provided the two three-spheres have the same radius.
Such a compactification was considered previously [25], with a different ansatz
for the dilaton field $\phi$, but the solution found there was unacceptable
[26]
(because singular in $\phi$). In fact, in [26] a `no-go' theorem was proved,
under certain premises, that rules out such compactifications. The
solution found here evades this theorem because the linear dilaton was
excluded by the premises of the theorem.

To determine how many supersymmetries are preserved by the $S^3$
compactification we need the fermion supersymmetry transformation laws in a
bosonic background. These are
$$
\eqalign{
\delta\psi_M &=\nabla_M\big(\omega^-\big)\epsilon \equiv \partial_m\epsilon +
\omega^-_{mAB}\Gamma^{AB}\epsilon \cr
\delta\lambda &= -{\sqrt{2}\over 4}\big(
\Gamma^M\partial_M\phi -{1\over12}\Gamma^{MNP}H_{MNP}\big)\epsilon }
\eqno (12)
$$
where $\omega^-$ is the connection with torsion $\omega^- =\omega -{1\over2}H$
(we use here the results of [27] with $\omega\to -\omega$ and a rescaled $H$).
We have $\delta\psi_M=0$ for our solution because $-{1\over2}H$ is the
parallelizing torsion for the 3-sphere, while $\delta\lambda=0$ implies, using
$k'=k$, that
$$
{1\over 2} n\cdot \Gamma \epsilon -{1\over12} \Gamma^{abc} H_{abc} \epsilon =0
\eqno (13)
$$
Since $\epsilon$ is chiral, $\Gamma^{abc}\epsilon
=\varepsilon^{abc}\gamma_7\epsilon$ where $\gamma_7=\Gamma^0\Gamma^1\cdots
\Gamma^6$, which satisfies $(\gamma_7)^2=1$. Hence (13) implies that
$$
\gamma_7\, n\cdot\Gamma \epsilon =\epsilon
\eqno (14)
$$
The matrix $\gamma_7\, n\cdot\Gamma$ squares to unity and has zero trace, which
means that of the 16 possible linearly independent chiral d=10 spinors only 8
linearly independent combinations satisfy (13). The solution therefore breaks
half the supersymmetry. Thus, the d=10 fivebrane, which itself breaks half the
supersymmetry, does not interpolate between {\sl maximally} supersymmetric
vacua.

Up to now we have regarded the d=10 fivebrane as a solution of N=1 d=10
supergravity, but it is also a solution of the d=10 N=2A and N=2B supergravity
theories. We conclude with some remarks on these cases. In the case of the
N=2A theory, one would expect, in view of the fact that this theory is the
dimensional reduction of d=11 supergravity, that the worldvolume field content
should be the same as in that theory, i.e. the six-dimensional N=4
antisymmetric tensor multiplet, which contains five worldvolume scalars. This
is
known to be true from an analysis of small fluctuations about the d=10 N=2A
fivebrane solution [5]. The d=10 N=2B theory is more problematic; a similar
small fluctuation analysis [5] led to the conclusion that the worldvolume
fields are those of the N=4 Maxwell supermultiplet. This does not fit easily
with the proposal that the linearized worldvolume field theory (at least) is to
be identified as a supersingleton field theory because the N=4 Maxwell theory
is
not conformally invariant in six dimensions. It may be that this is
attributable to the fact that the core of the d=10 fivebrane is not asymptotic
to $(adS)_7\times S^3$ but rather $\M_7\times S^3$, as we have seen, but it may
also be that the worldvolume field content has been mis-identified. Evidence
for mis-identification is the fact that the N=2B theory has a rigid $U(1)$
symmetry which is broken by the fivebrane solution; one would therefore expect
a
worldvolume Goldstone field for this broken $U(1)$, in addition to the four
Goldstone modes of broken translation invariance. This would mean that there
should be {\sl five} rather than four worldvolume scalars, just as for the N=2A
fivebrane. Given that the only worldvolume fermions are those arising from the
partial breaking of supersymmetry, these fields can only fit into an
antisymmetric tensor multiplet, as for the N=2A case.

\bigskip
\centerline{\bf Acknowledgements}
\medskip

PKT would like to thank Nathan Berkovits for helpful discussions.

\vfill\eject

\centerline{\bf References}

\item {[1]}
A. Linde, Phys. Lett. {\bf 129B} (1983) 177.

\item {[2]}
P. van Baal, F.A. Bais and P. van Nieuwenhuizen, Nucl. Phys. {\bf
B233} (1984) 477.

\item {[3]}
G.W. Gibbons, in {\it Supersymmetry, Supergravity and Related
Topics}, eds. F. del Aguila, J. A. de Azc{\' a}rraga and L.E. Iba{\~
n}ez, (World Scientific 1985).

\item {[4]}
G.W. Gibbons and K. Maeda, Nucl. Phys. {\bf B298} (1988) 741.

\item {[5]}
C. Callan, J. Harvey and A. Strominger, Nucl. Phys. {\bf B359} (1991)
611.

\item {[6]}
G. Horowitz and A. Strominger, Nucl. Phys. {\bf B360} (1991) 197.

\item {[7]}
M.J. Duff and X. Lu, Nucl. Phys. {\bf B354} (1991) 141; Phys. Lett. {\bf 273B}
(1991) 409.

\item {[8]}
M.J. Duff and K.S. Stelle, Phys. Lett. {\bf 253B} (1991) 113.

\item {[9]}
R. G{\" u}ven, Phys. Lett. {\bf 276B} (1992) 49.

\item {[10]}
P.G.O. Freund and M.A. Rubin, Phys. Lett. {\bf 97B} (1980) 233; M.J. Duff and
C.N. Pope, in {\it Supersymmetry and Supergravity '82}, eds. S. Ferrara, J.G.
Taylor and P. van Nieuwenhuizen (World Scientific, 1983); B. Biran, F.
Englert, B.de Wit and H. Nicolai, Phys. Lett. {\bf 124B} (1983) 45.

\item {[11]}
K. Pilch, P.K. Townsend and P. van Nieuwenhuizen,
Nucl. Phys. {\bf B242} (1984) 377.

\item {[12]}
J.H. Schwarz, Nucl. Phys. {\bf B226} (1983) 269.

\item {[13]}
R. Kallosh and A. Peet, Phys. Rev. {\bf D46} (1992) R5223.

\item {[14]}
M.J. Duff and X. Lu, Nucl. Phys. {\bf B390} (1993) 276.

\item {[15]}
M. G{\" u}naydin and N. Marcus, Class. Quantum Grav. {\bf 2} (1985)
L1; M. G{\" u}naydin P. van Nieuwenhuizen and N.P. Warner, Nucl. Phys.
{\bf B255} (1985) 63; H.S. Kim, L.J. Romans and P. van Nieuwenhuizen,
Phys. Rev. {\bf D32} (1985) 389.

\item {[16]}
C. Fronsdal, Phys. Rev. {\bf D26} (1982) 1988.

\item {[17]}
H. Nicolai and E. Sezgin, Phys. Lett. {\bf 143B} (1984) 389.

\item {[18]}
E. Bergshoeff, E. Sezgin and Y. Tanii, Int. J. Mod. Phys. A5 (1990) 3599; E.
Bergshoeff, A. Salam, E. Sezgin and Y. Tanii, Phys. Lett. {\bf 205B} (1988)
237.

\item {[19]}
H. Nicolai, E. Sezgin and Y. Tanii, Nucl. Phys. {\bf B305} [FS23] (1988) 483.

\item {[20]}
E. Bergshoeff, M.J. Duff, C.N. Pope and E. Sezgin, Phys. Lett. {\bf
199B} (1987) 69.

\item {[21]}
P.S. Howe, G. Sierra and P.K. Townsend, Nucl. Phys.  {\bf B221} (1983) 331.

\item {[22]}
M. G{\" u}naydin, B.E.W. Nilsson, G. Sierra and P.K. Townsend, Phys.
Lett. {\bf 176B} (1986) 45.

\item {[23]}
M.P. Blencowe and M.J. Duff, Phys. Lett. {\bf 203B} (1988) 229.

\item {[24]}
W.E. Couch and R.J. Torrence, GRG {\bf 16} (1984) 789.

\item {[25]}
M.J. Duff, P.K. Townsend and P. van Nieuwenhuizen, Phys. Lett. {\bf
122B} (1983) 232.

\item {[26]}
D.Z. Freedman, G.W. Gibbons and P.C. West, Phys. Lett. {\bf 124B} (1983) 491.

\item {[27]}
E. Bergshoeff and M. de Roo, Phys. Lett. {\bf 218B} (1989) 210.

\end